\documentclass{article}
\usepackage{cite}

\usepackage{arxiv}
\usepackage{subfig}
\usepackage{float}
\usepackage{textcomp}
\usepackage{graphicx}
\usepackage{amsmath,amssymb,amsfonts}
\usepackage{algorithmic}
\usepackage{graphicx}
\usepackage{xcolor}
\usepackage{array}
\usepackage[bottom]{footmisc}
\usepackage{multirow}

\usepackage[utf8]{inputenc} 
\usepackage[T1]{fontenc}    
\usepackage{hyperref}       
\usepackage{url}            
\usepackage{booktabs}       
\usepackage{amsfonts}       
\usepackage{nicefrac}       
\usepackage{microtype}      
\usepackage{lipsum}		
\usepackage{graphicx}
\usepackage{natbib}
\usepackage{doi}

\title{Smart OMVI: Obfuscated Malware Variant Identification using a novel dataset}

\author{ \href{https://orcid.org/0000-0001-6528-1681}{\includegraphics[scale=0.06]{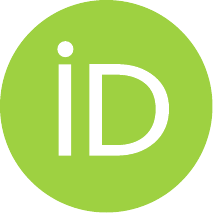}\hspace{1mm}Suleman Qamar}\thanks \\
	\\Department of Computer Science\\
	CIPMA Lab, PIEAS\\
	\texttt{m.sulemanqamar@gmail.com} \\
}

\renewcommand{\shorttitle}{\textit{Smart} OMVI}

\hypersetup{
pdftitle={Smart OMVI: Obfuscated Malware Variant Identification using a novel dataset},
pdfsubject={AI ML.},
pdfauthor={Suleman Qamar},
pdfkeywords={First keyword, Second keyword, More},
}

\newcolumntype{P}[1]{>{\centering\arraybackslash}p{#1}}
\newcolumntype{M}[1]{>{\centering\arraybackslash}m{#1}}
\def\BibTeX{{\rm B\kern-.05em{\sc i\kern-.025em b}\kern-.08em
    T\kern-.1667em\lower.7ex\hbox{E}\kern-.125emX}}
    
\begin{document}
\maketitle

\begin{abstract}
Cybersecurity has become a significant issue in the digital era as a result of the growth in everyday computer use. Cybercriminals now engage in more than virus distribution and computer hacking.
Cyberwarfare has developed as a result because it has become a threat to a nation's survival. Malware analysis serves as the first line of defence against an attack and is a significant component of cybercrime. Every day, malware attacks target a large number of computer users, businesses, and governmental agencies, causing billions of dollars in losses. Malware may evade multiple AV software with a very minor, cunning tweak made by its designers, despite the fact that security experts have a variety of tools at their disposal to identify it.
To address this challenge, a new dataset called the Obfuscated Malware Dataset (OMD) has been developed. This dataset comprises 40 distinct malware families having 21924 samples, and it incorporates obfuscation techniques that mimic the strategies employed by malware creators to make their malware variations different from the original samples. The purpose of this dataset is to provide a more realistic and representative environment for evaluating the effectiveness of malware analysis techniques.
Different conventional machine learning algorithms including but not limited to Support Vector Machine (SVM), Random Forrest (RF), Extreme Gradient Boosting (XGBOOST) etc are applied and contrasted. The results demonstrated that XGBoost outperformed the other algorithms, achieving an accuracy of f 82\%, precision of 88\%, recall of 80\%, and an F1-Score of 83\%.
\end{abstract}


\section{Introduction}
Malicious software shortened to malware, is a piece of software made with the intention of breaking into and causing harm to computers without the user's knowledge. The word "malware" refers to a broad category of destructive software; some of the most popular varieties are listed in table 1.
Software malware may come in a variety of shapes and sizes. Desktops, servers, mobile phones, printers, and programmable electrical circuits are just a few possible deployment platforms. Sophisticated assaults have proven that data may be taken using well-written malware that only exists in system memory and leaves no trace in the form of permanent data. Information security safeguards like desktop firewalls and anti-virus software have been reported to be disabled by malware. Some are even capable of compromising audit, authentication, and authorisation processes.
\\
\begin{table*}
\caption{Common Malware Types}
\label{tab:audit_logs}
\setlength{\tabcolsep}{3pt}
\begin{tabular}{|M{75pt}|M{400pt}|}
\hline
 \textbf{Type} & \textbf{Description} \\ 
  \hline
 Adware \cite{gao2019should} & They display unwanted advertisements on a computer or mobile device resulting in slowdowm. It often gets installed alongside other software without the user's knowledge or consent, and can collect data about the user's browsing habits.   \\  
   \hline
 Bot & Automated programs that can perform various tasks, including useful and malicious ones, on the internet. Botnets, networks of infected computers controlled by a botmaster, can be used to carry out large-scale attacks.  \\   
   \hline
 Bug & Flaws or errors in software code that can cause unexpected behavior ranging from minor glitches to serious security vulnerabilities that can be exploited by attackers to gain unauthorized access or cause system crashes. \\  \hline
 Ransomware & A type of malware that encrypts a victim's files and demands payment in exchange for the decryption key. These attacks can be devastating for individuals and organizations, causing loss of data and significant financial costs. \\  \hline
 Rootkit & It allows an attacker to gain root-level access to a computer system which can be used to hide other malware, steal sensitive data, or control the system remotely. Usually, rootkits are very difficult to detect and remove, often requiring specialized tools and expertise. \\
  \hline
  Spyware & A type of malicious software designed to gather sensitive information such as personal information including but not limited to passwords, credit card numbers, and browsing history from a computer or mobile device \\ \hline
  Trojan Horse & Disguised as a legitimate program, but once installed, can give unauthorized access to a computer system or steal sensitive data. A variety of harmful actions, such as deleting files, stealing passwords, or opening up a backdoor for remote access can be performed by them. \\ \hline
  Virus & Self-replication and fast spread are the identifying features of a virus. These can cause a range of damages, from data loss to system crashes, and can be difficult to remove once activated.\\ \hline
  Worms & These can cause harm by overloading systems, stealing information, or carrying out other malicious actions. Some famous examples of worms include Code Red, Conficker, and WannaCry.\\ \hline
  Scareware & A type of malware that tries to trick users into thinking their computer is infected with a virus or other threat. These typically displays fake pop-ups or alerts urging the user to purchase bogus security software or services which can be harmful if users fall for the scam and download the fake software, which may contain actual malware.\\ \hline
  Fileless Malware & A type of malware that operates entirely in a device's memory or system registry, leaving little to no trace on disk and thus, is harder to detect and remove than traditional malware because it doesn't leave the same types of artifacts. Examples of fileless malware include PowerShell-based attacks and memory-based exploits.\\ \hline
  Keyloggers & These capture keystrokes typed on a device's keyboard, potentially allowing the attacker to steal sensitive information like passwords or credit card numbers. They can be either software or hardware-based, with the latter being more difficult to detect. Some even have the ability to capture screenshots or record audio or video from the device.\\ \hline
  Wiper & A type of malware that is designed to destroy or wipe out data on a device or network. These attacks can be devastating, as they can render a system completely unusable and may be difficult or impossible to recover from. Some notable examples of wiper malware include Shamoon, NotPetya, and OlympicDestroyer.\\ \hline
\end{tabular}
  \label{tab:WinLogs}
\end{table*}
Even when a compromised machine is rebooted, startup files have been set to preserve persistence. When run, advanced malware may duplicate itself or remain dormant until called upon by its command features to extract data or delete files. Four operational characteristics often serve to characterise a particular piece of malware:
\begin{enumerate}
    \item Propagation: The method through which malware spreads across several systems.
    \item Infection: The malware's method of installation and its capacity to withstand cleanup efforts after being set up.
    \item Self-Defense: The technique utilised to obfuscate its existence and thwart examination.
    \item Capabilities: Functions accessible by malware operator.
\end{enumerate}
These might also be referred to as anti-reversing capabilities.
\begin{table}
\caption{List of Abbreviations}
\centering
\label{tab:abbr}
\begin{tabular}{|M{75pt}|M{143pt}|}
\hline
\textbf{Abbreviations}&\textbf{Full Form} \\
\hline
OMD & 
Obfucated Malware Datase \\\hline
RNN  & Recurrent Neural Network \\\hline
PUA & Potentially Unwanted Applications \\\hline
CCV  & Card Code Verification \\\hline
ASM  & Assembly language\\\hline
NOP & No Operation\\ \hline
\end{tabular}
\end{table}
\subsection{Different Families of Malware}
Malware are categorized  into different families on the basis of their behavior and type of damage done to host machine.

\subsection{Criminal Steps of Malware}
Malware are diverse in their kind and activity, attacking different types of information in order to cause problems for the user. 
\begin{enumerate}
\item Intelligence gathering: A criminal searches the target for weak spots in order to prepare an assault.
\item Preparation: A criminal develops, tweaks, or somehow acquires malware to meet the demands of an attack.
\item Distribution: The spread of malware takes place.
\item Compromise: Malware infects the system.
\item Demand: The power of malware is released.
\item Execution: Malware transfers data to the malware operator, a process known as exfiltration, which achieves the attack's goal by transferring data from an information system without any type of consent.
\end{enumerate}
Modern malware remediation is getting harder and harder to do for a variety of reasons. Malware that exploits the use zero-day vulnerabilities \cite{kumar2022zero, barros2022malware} comes in a much wider range of forms. A vulnerability is described as "zero-day" if potential victims are unaware of it, in which case they have no time to prepare. Malware is now capable of taking on several forms thanks to polymorphic design. Malware that is polymorphic alters some aspects of itself with each infection. This modification may take the form of an unusable code update. This method avoids signature-based detection \cite{botacin2022heaven} methods since they frequently create a distinct signature from a file containing malware using a hash algorithm, meaning that any modification to the file would alter its signature. Moreover, because polymorphic \cite{selamat2022polymorphic} malware has the ability to alter its own filename upon infection, standard signature-based methods of detection are hampered. As a result, to address this problem, a new dataset called the Obfucated Malware Dataset (OMD) is presented.
The following are the overall contributions of this work:
\begin{enumerate}
 \item	A large malware dataset named Obfucated Malware Dataset (OMD) is generated by collecting malware data from different sources, combining them with two other datasets, namely, Malimg and Kaggle’s Microsoft
Malware Classification Challenge (BIG 2015) Dataset and applying various obfuscation techniques resulting in 40 different families of malware. The final dataset contains 40 classes and 21924 samples. 
 \item In contrast to other datasets, all samples in OMD are obfuscated using different techniques resulting in a dataset that can mimic new or polymorphic malwares. Traditional machine learning techniques are applied on the dataset and results are compared and contrasted.
 \end{enumerate}
 The rest of the paper is organized as follows. The next section highlights the related work in the
field of malware classification. Section 3 explains our novel classification framework
methodology, while section 4 discusses the experimental setup. Section 5 presents the result
analysis and discussion of our work. Section 6 concludes the paper.

\section{Related Work}
Any software that is intentionally designed to disrupt a computer, server, client, or computer network, leak sensitive information, allow unauthorised access to data or systems, deny access to information, or inadvertently compromise user privacy and security on computers is known as malware, a combination of words malicious and software \cite{brewer2016ransomware, tahir2018study}.
\begin{table}
\caption{Stolen data prices}
\label{mal_price}
\centering
\setlength{\tabcolsep}{3pt}
\begin{tabular}{|l|l|}
\hline
\textbf{Data type}&\textbf{Price} \\\hline
CCV& \$3.25 \\\hline
OS administrative login&  \$2.50 \\\hline
FTP exploit& \$6.00 \\\hline
Full identity information & \$5.00\\\hline
Rich bank account credentials & \$750.00\\ \hline
US passport information & \$800.00 \\ \hline
Router credentials &\$12.50 \\ \hline
\end{tabular}
\end{table}

\begin{table}
\caption{Cost of Malware and Crimeware}
\label{mal_price2}
\centering
\setlength{\tabcolsep}{3pt}
\begin{tabular}{|l|l|}
\hline
\textbf{Theft Enabling Commodity}&\textbf{Price} \\\hline
Keystroke logger &\$25 on average \\\hline
Botnets & \$100 to \$200 per 1,000 infections, depending on location \\\hline
Spamming email service & \$.01 per 1,000 emails, reliability of more than 85\% delivered \\\hline
Shop admins (Credit Card databases) & \$100 to \$300\\\hline
Credit Card numbers without CCV2 & \$1 to \$3\\\hline
Credit Card numbers with CCV2 &\$1.50 to \$10.00, depending on the country\\\hline
Socks accounts & \$5 to \$40/month\\\hline
Sniffer dumps &\$50 to \$100/month\\\hline
Western Union exploits &\$300 to \$1,000\\\hline
Remote desktops &\$5 to \$8\\\hline
Scam letters &\$3 to \$5\\\hline
\end{tabular}
\end{table}
The underground market for stolen data is the chief reason of malware. In several forums, data hackers may resell their loot \cite{menn2010fatal}. Table \ref{mal_price} \cite{geer20090wned} contains examples of prices paid for various categories of stolen data. The money demanded for the stolen goods indicated in Table \ref{mal_price} drove the development of secondary malware marketplaces, which result in software tools that make malware more and more successful at facilitating information theft. In general, people utilise software to automate laborious and resource-intensive jobs, and malware authors are no different. Automating the transmission of malware and the data collection process lowers operating expenses while enabling criminals to conceal their operations. Systems for the distribution and operation of malware have gotten more and more modular.
Crimeware \cite{salloum2022systematic, wang2021deep} is a term used to describe software that has been found to have such malware support systems. The Zeus toolkit \cite{grammatikakis2021understanding} is a good illustration of crimeware. The Zeus virus first appeared in 2006, and the associated crimeware in 2007. Zeus' crimeware takes use of its modular nature, so attackers may modify and deploy new capabilities relatively rapidly. An attacker may choose the features to be included in a "release" and a unique encryption key for data that has been captured using an intuitive graphical interface. Zeus crimeware has been used to produce more than 5,000 different versions of the Zeus software. Although numerous Zeus users have been identified and prosecuted with cybercrimes, the Zeus crimeware writers remain at large \cite{kazi2022comparing, rose2022ideres}.
\subsection{Detecttion and Classfication}
For the purpose of classifying and detecting malware, both static and dynamic analysis approaches have been widely used. The development of machine learning has created a wide range of possibilities for analysis and forecasting for both malware analysis methods. Visual malware image-based categorization is a relatively new development in the field of malware analysis. The textural elements of the malware's visual image file were discovered in 2011 by Nataraj et al. \cite{nataraj2011malware} . These files are generated by translating the byte code of a portable executable (PE) binary file to the pixel value's grey level. They used wavelet decomposition to obtain the textural characteristics from the malware picture. The K-nearest Neighbor machine learning algorithm is then used on these characteristics. 

\cite{gibert2019using} suggested a simple design for a convolutional neural network made up of three convolutional blocks, one fully-connected block, and one output layer. ReLU activation, max-pooling, normalization, and a convolution operation made up each convolution block. The convolutional layers served as detection filters for certain features or patterns in the input, while the following fully-connected layers combined the learnt information to produce a particular target output. The effectiveness of their method was tested on the Microsoft Malware Classification Challenge \cite{ronen2018microsoft}  versus manually created feature extractors \cite{kancherla2013image, ahmadi2016novel}, and the findings show that deep learning architectures perform better at identifying malware represented as grayscale photos. Similar to this, \cite{rezende2017malicious} performed classification on the MalImg \cite{nataraj2011malware} dataset using the ResNet-50 architecture with pretrained weights.
\subsection{Notable Datasets}
Malware datasets with coarse family labels are shown in Table \ref{Notable_Datasets}. The MalImg, VX Heaven \cite{qiao2016automatically}, Kaggle, and MalDozer \cite{karbab2018maldozer} datasets' collecting periods are unrecorded; publishing dates are taken as an upper limit for the period's conclusion. Drebin's \cite{arp2014drebin} labels appear to have been combined from those of 10 other antivirus programs, while the precise labelling process is unknown. The Microsoft Security Essentials program was used to label the MalImg dataset. The VX Heaven website was active from 1999 to 2012, and the malware in the collection is thought to be extremely old. The Kaspersky antivirus software was used to label the VX Heaven dataset.
MalDozer's labelling strategy was not made public, however family names imply that one antivirus was used. Family labels are not present in the initial EMBER \cite{anderson2018ember} dataset, but an extra 1,000,000 files—both harmful and benign—were made available in 2018. AVClass \cite{sebastian2016avclass} labels indicate that 485,000 of these files are malware samples.  The Malpedia \cite{plohmann2017malpedia} collection includes labels that were received from open-source reporting, and some malware samples were dumped and unpacked using human analysis. Other family designations, however, were generated automatically using tools like YARA rules and comparisons of unpacked files to known malware samples.

 The bulk of files in the Malsign \cite{kotzias2015certified} collection are not malicious programs but rather PUAs. Malsign reference labels were created by clustering characteristics that had been statically extracted. MaLabel has 115,157 samples, of which 46,157 are part of 11 major families and the rest 69,000 are a part of families with fewer than 1,000 samples. The dataset contains an unknown number of families in total. Microsoft provided a collection of 1.3 million malware samples, labelled using a combination of antivirus labelling and manual labelling, to the developers of the MtNet \cite{huang2016mtnet} malware classifier.

Although there are some datasets that has a few obfuscated malware samples, no dataset is purely focused on obfuscated malware classification. Using malware reference datasets with these proprties may yield evaluation results that are biased or incorrect for newer malwares. There aren't many prominent datasets that contain malware that targets other operating systems (including Linux, macOS, and iOS), but this research is outside the purview of our article.

\begin{table*}
\caption{Notable Datasets}
\label{Notable_Datasets}
\setlength{\tabcolsep}{3pt}
\begin{tabular}{|M{69pt}|M{50pt}|M{60pt}|M{45pt}|M{63pt}|M{83pt}|M{86pt}|}
\hline
Name  & Year& Samples & Family & Operating System & Labelling Methodology  & Period of collection\\\hline
MOTIF&2022& 3,095&454& Windows&  Threat Reports&Jan. 2016 - Jan. 2021\\\hline
MalImg& 2011 & 9,458 &25& Windows &Single AV& July 2011 or earlier\\\hline
Kaggle&  2018&10,868 &9& Windows& Susp. Single AV& Feb. 2015 or earlier\\\hline
AMD&2017& 24,553 &71 &Android  &Cluster Labeling&2010 - 2016\\\hline
MalDozer &2018&20,089 &32& Android&Susp. Single AV& Mar. 2018 or earlier \\\hline
EMBER&2018 &485,000 &3,226& Windows& AVClass& 2018 \\\hline
MalGenome &2015 &1,260 &49 &Android  &Threat Reports&Aug. 2010 - Oct. 2011\\\hline
Variant &2015 &85 &8 &Windows &Threat Reports&Jan. 2014 \\\hline
Malheur Rieck&2006& 3,133& 24& Windows  &AV Majority Vote&2006 - 2009\\\hline
Drebin &2010&5,560& 179 &Android & AV-based&Aug. 2010 - Oct. 2012\\\hline
VX Heaven &2016& 271,092 &137& Windows &Single AV&2012 \\\hline
Malicia &2012 &11,363 &55 &Windows &Cluster Labeling&Mar. 2012 - Mar. 2013 \\\hline
Malpedia &2017& 5,862 &2,165 &Both   &Hybrid&2017- ongoing\\\hline
Malsign &2015& 142,513 &Unknown &Windows   &Cluster labeling&2012- 2014\\\hline
MaLabel &2015& 115,157 &>80 &Windows  &AV Majority Vote&Apr. 2015\\\hline
MtNet &2016& 1,300,000 &98 &Windows  &Hybrid&Jun. 2016\\\hline
\end{tabular}
\end{table*}
\section{Methodology}
\begin{table}
\caption{Three datasets used to create OMD}
\label{Three_datasets}
\centering
\setlength{\tabcolsep}{3pt}
\begin{tabular}{|M{150pt}|M{70pt}|M{70pt}|}
\hline
\textbf{Dataset Name} & \textbf{Number of Families} & \textbf{Number of Samples} \\\hline
Malimg   &25 &9339\\\hline
Kaggle's Microsoft Malware Classification Challenge (BIG 2015) & 9&	10868\\\hline
TinyOMD&	6& 489\\\hline
Total	&40& 21924\\\hline
\end{tabular}
\end{table}
\begin{table}
\caption{Malimg Dataset}
\label{Malimg_Dataset}
\centering
\setlength{\tabcolsep}{3pt}
\begin{tabular}{|M{25pt}|M{80pt}|M{65pt}|M{65pt}|}
\hline
S.No.&Family&Family Name & Number of samples \\
\hline
$1$ &	Dailer &	Adialer.C &	122 \\\hline
$2 $&	Backdoor &	Agent.FYI &	116\\\hline
$3 $&	Worm&	Allaple.A	&2949\\\hline
$4 $&	Worm&	Allaple.L	&1591\\\hline
$5 $&	Trojan&	Alueron.gen!J&	198\\\hline
$6 $&	Worm:AutoIT&	Autorun.K&	106\\\hline
$7 $&	Trojan	&C2lop.gen!G&	200\\\hline
$8$&	Trojan	&C2lop.p	&146\\\hline
$9$&	Dailer	&Diaplaform.B&	177\\\hline
$10$ &	TrojanDownloader&	Dontovo.A&	162\\\hline
$11 $&	Rogue	&Fakerean	&381\\\hline
$12 $&	Dailer	&Instantaccess	&431\\\hline
$13 $&	PWS&	Lolyda.AA1	&213\\\hline
$14 $&	PWS&	Lolyda.AA2	&184\\\hline
$15 $&	PWS&	Lolyda.AA3	&123\\\hline
$16 $&	PWS&	Lolyda.AT	&159\\\hline
$17 $&	Trojan&	Malex.gen!J	&136\\\hline
$18 $&	TrojanDownloader&	Obfuscated.AD	&142\\\hline
$19 $&	Backdoor&	Rbot!gen	&158\\\hline
$20 $&	Trojan&	Skintrim.N	&80\\\hline
$21 $&	TrojanDownloader&	Swizzor.gen!E&	128\\\hline
$22$ &	TrojanDownloader&	Swizzor.gen!I&	132\\\hline
$23 $&	Worm&	VB.AT	&408\\\hline
$24 $&	TrojanDownloader&	Wintrim.BX&	97\\\hline
$25 $&	Worm&	Yuner.A	&800\\\hline
& \textbf{Total} & &\textbf{9339}\\ \hline
\end{tabular}
\end{table}
\begin{table}
\caption{Obfuscated Malware Dataset (OMD))}
\label{MicMal}
\centering
\setlength{\tabcolsep}{3pt}
\begin{tabular}{|M{25pt}|M{80pt}|M{65pt}|M{65pt}|}
\hline
S.No.&Family&Family Name & Number of samples \\
\hline
$1$&	Backdoor&	Gatak	&1013\\\hline
$2$ &	Backdoor&	Kelihos\_ver1	&398\\\hline
$3$&	Backdoor&	Kelihos\_ver3	&2493\\\hline
$4 $&	Adware	&Lollipop	&2476\\\hline
$5 $&	Any Obfuscated Malware&	Obfuscator.ACY	&1228\\\hline
$6 $&	Worm&	Ramnit	&1541\\\hline
$7 $&	Backdoor&	Simda	&42\\\hline
$8$ &	TrojanDownloader&	Tracur	&751\\\hline
$9$&	Trojan&	Vundo	&475\\\hline
& \textbf{Total} & &\textbf{10868}\\ \hline
\end{tabular}
\end{table}

\subsection{Dataset Generation}
\begin{figure}\centering
\subfloat{{\includegraphics[width=9cm]{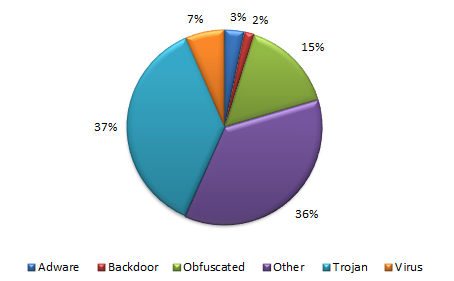} }}%
    \caption{TinyOMD Classes with percentage of samples.}
    \label{TinyOMD}
    \end{figure}

\begin{figure*}\centering
\subfloat{{\includegraphics[width=18cm]{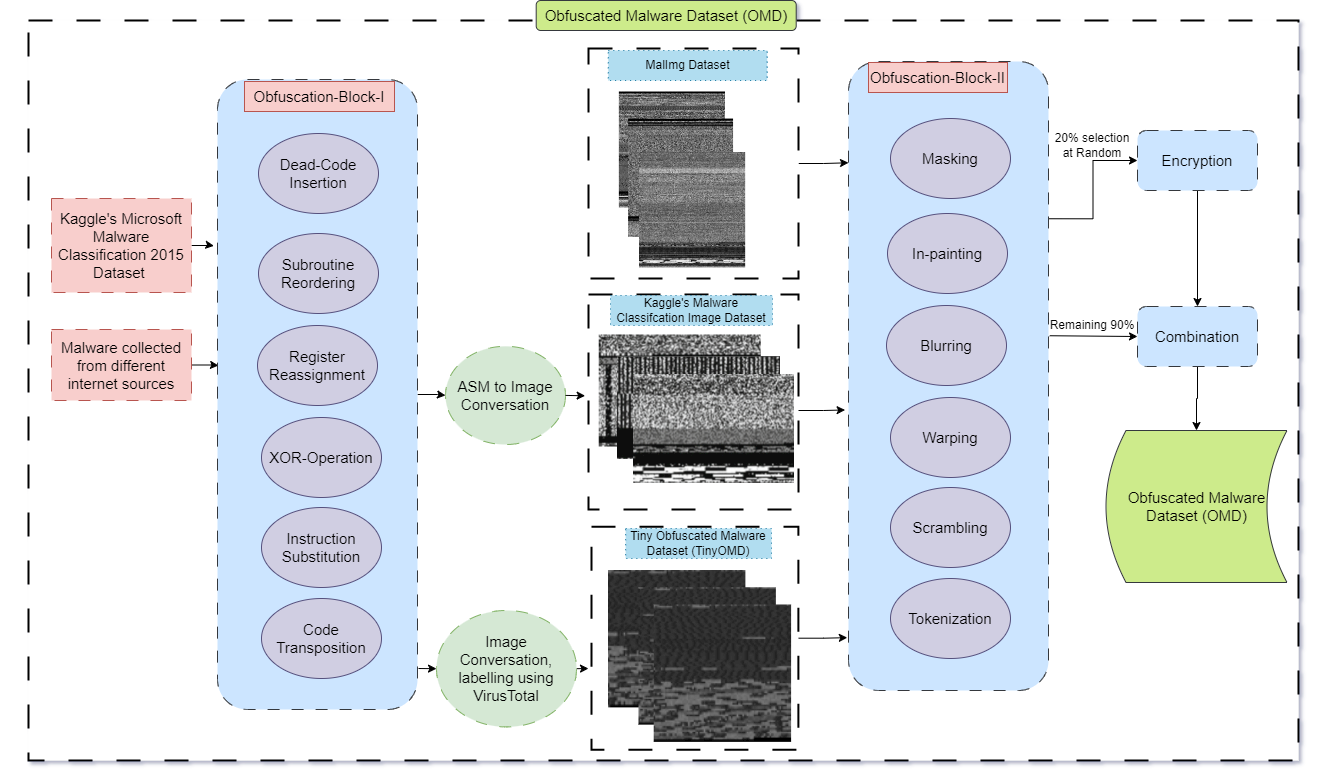} }}%
    \caption{Dataset Generation Workflow}
    \label{workflow}
    \end{figure*}

Dataset is one of the most important things in machine and deep learning. An appropriate dataset was required that mimics the polymorphism of modern malware families, hence, a dataset named as OMD shown in fig. \ref{workflow} is created using three smaller datasets given in table \ref{Three_datasets}. First dataset is Malimg Dataset represented in table \ref{Malimg_Dataset} that comprises of 9339 malware samples which were classified into 25 malware families. This dataset contains grey-scale images formed from malware binaries. A 2D matrix was generated from these malware binaries and then represented as grey-scale images.  Second dataset was Kaggle Malware Classification challenge 2015 dataset represented in table \ref{MicMal} containing 10868 malware samples that contain 9 different malware families. This dataset doesn't contain malware samples in the form of grey-scale images, instead, it has ASM and Byte type malware sample files. ASM files were taken, different obfuscations were applied, and then these files were converted to grey-scale in order to add them to the dataset.
Third dataset \ref{OMD} was generated using malware samples collected from different sources and manually labelling them with the help of VirusTotal.
Similar obfuscation techniques were also applied on this dataset. It was named as Tiny Obfuscated Malware Dataset (TinyOMD) shown in figure \ref{TinyOMD} and has 489 samples representing 6 classes given in Table \ref{OMD}.
So, to Conclude three dataset namely Malimg, Kaggle Malware Classification challenge 2015 dataset and TinyOMD amounting to a total of 20696 and 40 classes given in table \ref{Three_datasets}.
Next step is Obfuscation, multiple techniques are applied for data obfuscation which explained in obfuscation section. 
\subsection{Obfuscation}
Obfuscation is applied by help of two obfuscation blocks each containing six different obfuscation techniques, furthermore, encryption is applied to 20\% of malware samples at random and combined with the reaming 80\% to formulate the Obfuscated Malware Classification Dataset. 
\subsubsection{Obfuscation Block-I} Six obfuscation techniques applied in the first block were Dead-Code Insertion, Subroutine Reordering, Register Reassignment, XOR-Operation, Instruction Substitution and Code Transposition. This block is applied on ASM sample files in Kaggle's Microsoft Classification 2015 Dataset and the newly created Tiny Obfuscated Malware Dataset (TinyOMD). All these techniques change malware source code thus resulting in change of signature without having any impact on its functionality.
Dead-Code Insertion also known as NOp-Insertion is an instruction that itself doesn't result in any sort of change in the functionally of the malware but is used to change its signature. It is commonly known as dead obfuscation.
Code Transposition also known as Jump instruction transfers the program sequence to the memory address given in the operand based on the specified flag. This results in complex and difficult to understand code obfuscation. In Register Reassignment, extra lines of code are added, again resulting in change in malware signature.
\subsubsection{Obfuscation Block-II}
 Six different obfuscation techniques were also applied in the second obfuscation block, namely, masking, in-painting, blurring, warping, scrambling, and tokenization. This block is applied on image sample files in all three sub-datasets including Malimg dataset, Kaggle's Microsoft Classification 2015 Dataset and the newly created Tiny Obfuscated Malware Dataset (TinyOMD). These techniques make malware detection even more difficult. 
Furthermore, around 10\% malware samples were taken at random from overall dataset, encrypted and added back to the dataset.

\begin{figure}\centering
\subfloat{{\includegraphics[width=9cm]{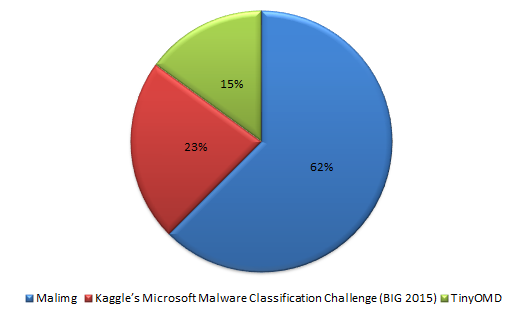} }}%
    \caption{Percentage of each dataset samples in OMD}
    \label{POMD}
    \end{figure}
\begin{figure}\centering
\subfloat{{\includegraphics[width=9cm]{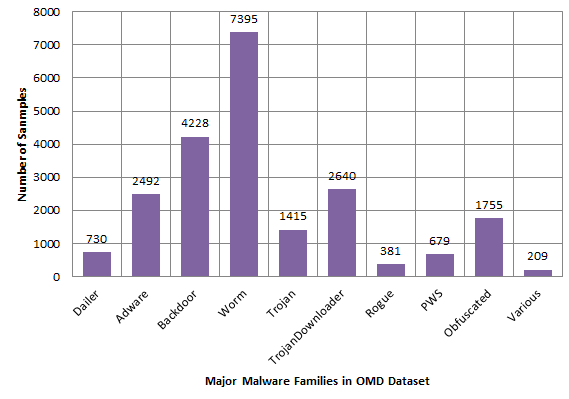} }}%
    \caption{Major malware families sample count in OMD}
    \label{OMDfam}
    \end{figure}
\begin{table}
\caption{Tiny Obfuscated Malware Dataset (TinyOMD)}
\label{tiny_OMD}
\centering
\setlength{\tabcolsep}{3pt}
\begin{tabular}{|M{75pt}|M{143pt}|}
\hline
\textbf{Class}&\textbf{Number of Samples} \\\hline
Adware	&16\\\hline
Backdoor&	8\\\hline
Obfuscated&	76\\\hline
Other	&177\\\hline
Trojan	&180\\\hline
Virus	&32\\\hline
Total	&489\\\hline
\end{tabular}
\end{table}

\begin{table}
\caption{Obfuscated Malware Dataset (OMD))}
\label{OMD}
\centering
\setlength{\tabcolsep}{3pt}
\begin{tabular}{|M{25pt}|M{100pt}|M{65pt}|M{65pt}|}
\hline
S.No.&Family&Family Name & Number of samples \\
\hline
$1$ &	Dailer &	Adialer.C &	122 \\\hline
$2 $&	Adware&	Various	& 16\\\hline
$3 $&	Backdoor &	Agent.FYI &	116\\\hline
$4 $&	Worm&	Allaple.A	&2949\\\hline
$5 $&	Worm&	Allaple.L	&1591\\\hline
$6 $&	Trojan&	Alueron.gen!J&	198\\\hline
$7 $&	Worm:AutoIT&	Autorun.K&	106\\\hline
$8 $&	Backdoor&	Various	&8\\\hline
$9 $&	Trojan	&C2lop.gen!G&	200\\\hline
$10$&	Trojan	&C2lop.p	&146\\\hline
$11$&	Dailer	&Diaplaform.B&	177\\\hline
$12$ &	TrojanDownloader&	Dontovo.A&	162\\\hline
$13 $&	Rogue	&Fakerean	&381\\\hline
$14 $&	Backdoor&	Gatak	&1013\\\hline
$15 $&	Dailer	&Instantaccess	&431\\\hline
$16$ &	Backdoor&	Kelihos\_ver1	&398\\\hline
$17 $&	Backdoor&	Kelihos\_ver3	&2493\\\hline
$18 $&	Adware	&Lollipop	&2476\\\hline
$19 $&	PWS&	Lolyda.AA1	&213\\\hline
$20 $&	PWS&	Lolyda.AA2	&184\\\hline
$21 $&	PWS&	Lolyda.AA3	&123\\\hline
$22 $&	PWS&	Lolyda.AT	&159\\\hline
$23 $&	Trojan&	Malex.gen!J	&136\\\hline
$24 $&	Obfuscated&	Various	&76\\\hline
$25 $&	TrojanDownloader&	Obfuscated.AD	&1370\\\hline
$26 $&	Any Obfuscated Malware&	Obfuscator.ACY	&1679\\\hline
$27 $&	Various&	Various		&177\\\hline
$28 $&	Worm&	Ramnit	&1541\\\hline
$29 $&	Backdoor&	Rbot!gen	&158\\\hline
$30 $&	Backdoor&	Simda	&42\\\hline
$31 $&	Trojan&	Skintrim.N	&80\\\hline
$32 $&	TrojanDownloader&	Swizzor.gen!E&	128\\\hline
$33$ &	TrojanDownloader&	Swizzor.gen!I&	132\\\hline
$34$ &	TrojanDownloader&	Tracur	&751\\\hline
$35 $&	Trojan&	Various	&180\\\hline
$36 $&	Worm&	VB.AT	&408\\\hline
$37 $&	Virus&	Various	&32\\\hline
$38 $&	Trojan&	Vundo	&475\\\hline
$39 $&	TrojanDownloader&	Wintrim.BX&	97\\\hline
$40 $&	Worm&	Yuner.A	&800\\\hline
& \textbf{Total} & &\textbf{21924}\\ \hline
\end{tabular}
\end{table}
\subsubsection{Data augmentation}
A lack of data causes deep learning models to overfit. To achieve effective generalisation and practical training, a large amount of data is therefore needed. Data augmentation is the process of increasing data samples by supplementing the underlying data \cite{shorten2019survey, naveed2021survey}. To enhance generality and make the suggested classification framework resistant to different types of malware data, we have expanded the training dataset in this study. Hence, it makes the suggested categorization system useful for categorising malware families. As illustrated in Table \ref{Augmentation}, the adopted augmentation technique incorporates a number of transformations, including reflections, scaling, rotation, and shear.
To increase the generality of the models, all trainings are done after data augmentation.
\begin{table}
\caption{Augmentation parameters}
\label{Augmentation}
\centering
\setlength{\tabcolsep}{3pt}
\begin{tabular}{|M{25pt}|M{95pt}|M{95pt}|}
\hline
S.No.&Augmentation type&Parameter \\\hline
$1$ &Rotate & [0, 360] degrees \\\hline
$2$&Shear &[-0.1,0.1]\\\hline
$3$&Reflection& X: [-1, 1],Y: [-1, 1]\\\hline
$4$&Scale &[0.2, 1]\\\hline
$5$&Horizontal Flip &-\\\hline
$6$&Vertical Flip &-\\\hline
$7$&Width Shift &[0, 0.2]\\\hline
$8$&Height Shift &[0, 0.2]\\\hline
\end{tabular}
\end{table}
\subsubsection{Dataset Partitioning}
In this study, the training and testing stages of the dataset partitioning strategy were 70-30.
The literature often mentions dataset partitioning as 80-20, 75-25, or even 70-30. Hence, 70-30 was choosen to make the model robust. In this context, the dataset's dimensions are important.
\section{Performance Metrics}
Prior to delving deeper into the performance measures, it was necessary to establish several fundamental units or classification categories, including TruePositive, TrueNegative, FalsePositive, and FalseNegative. To comprehend how each of these units is to be categorised, refer to Table .
\begin{enumerate}
    \item TruePositive (TP): Both the actual and anticipated labels for the data sample are positive. If the model predicts class name of sample correctly out of 40 class names then it will be considered a true positive.  

    \item TrueNegative (TN): In multi-class classification, true negative is sum of all classes except for the class which is being under consideration. 

	\item FalsePositive (FP): False positive will represent sum of all classes except true positive in the corresponding columns. 

    \item FalseNegative (FN): Similarly, false negative will represent sum of all classes except true positive in the corresponding rows.  
\subsection{Recall}
In multiclass classification, recall (also known as sensitivity or true positive rate) given in Eg. \ref{eqn:recall}  is a metric that measures the proportion of true positive predictions for a given class out of all the actual positive instances in that class. It is defined  as:
\begin{equation}
\label{eqn:recall}
Recall=\dfrac{True Positive}{True Positive+False Negative}    
\end{equation}
where True positives are the number of correctly classified instances of a specific class, and False negatives are the number of instances that belong to that class but are incorrectly classified as belonging to a different class.

In other words, recall in multiclass classification tells us how well the model is able to correctly identify all instances of a specific class, regardless of whether it misclassifies some instances from other classes as belonging to that class. A high recall value for a specific class indicates that the model is good at correctly identifying all instances of that class, while a low recall value indicates that the model is missing many instances of that class.

\subsection{Specificity}
Specificity is recall’s inverse, that is, it indicates how well a model is performing to correctly identify the negative labels. In simple words, specificity would be the ratio of TrueNegative to Total Negatives. Negative labels in our data would be the logs generated by benign applications. Specificity is calculated using formula given in Eq. \ref{eqn:Specificity}. 
\begin{equation}
\label{eqn:Specificity}
Specificity=\dfrac{True Negative}{True Negative+False Positive}    
\end{equation}
\subsection{Precision}
In multiclass classification, precision given in Eg. \ref{eqn:Precision} is a metric that measures the proportion of true positive predictions for a given class out of all the positive predictions made by the model for that class. It is defined as:
\begin{equation}
\label{eqn:Precision}
Precision=\dfrac{True Positive}{True Positive+False Positive}    
\end{equation}
where True positives are the number of correctly classified instances of a specific class, and False positives are the number of instances that are incorrectly classified as belonging to that class, when in fact they belong to a different class.

In other words, precision in multiclass classification tells us how well the model is able to correctly classify instances of a specific class, without misclassifying instances from other classes as belonging to that class. A high precision value for a specific class indicates that the model is good at correctly identifying instances of that class, while a low precision value indicates that the model is misclassifying many instances from other classes as belonging to that class.

\subsection{Accuracy}
Accuracy is how well the model is performing in correctly predicting the Positive and Negative Labels. It is the ratio correctly predicted to total samples. Eq. \ref{eqn:accuracy} is used to calculate the accuracy of any model. 
\begin{equation}
\label{eqn:accuracy}
Accuracy=\dfrac{True Positive+True Negative}{Total number of samples}    
\end{equation}
\subsection{F - Score}
F-score is also known as harmonic mean of precision and recall. This performance metric is highly suitable when it comes to imbalanced datasets. F-Score can be calculated using the model’s precision and recall given in \ref{eqn:fscore}.
\begin{equation}
\label{eqn:fscore}
F-Score=2*\dfrac{precision*recall}{precision+recall}
\end{equation}
\end{enumerate}
Performance metrics are summarized in table \ref{Performance_metrics}.
\begin{table}
\caption{Summarized Performance Metrics}
\label{Performance_metrics}
\centering
\setlength{\tabcolsep}{3pt}
\begin{tabular}{|M{50pt}|M{180pt}|}
\hline
Name & Formula \\
\hline
Accuracy& \begin{equation*}
\label{eqn:accuracy}
\dfrac{True Positive+True Negative}{Total number of samples}    
\end{equation*}\\\hline
Recall &	\begin{equation*}
\label{eqn:recall}
\dfrac{True Positive}{True Positive+False Negative}    
\end{equation*}\\\hline
Precesion & \begin{equation*}
\label{eqn:Precision}
\dfrac{True Positive}{True Positive+False Positive}    
\end{equation*}\\\hline
F1-Score &	\begin{equation*}
\label{eqn:fscore}
2*\dfrac{precision*recall}{precision+recall}
\end{equation*}
\\\hline
\end{tabular}
\end{table}
\section{Traditional Machine Learning Classifiers}
\subsection{Decision Tree}
Decision trees (DT) \cite{charbuty2021classification} are a non-parametric supervised leaning approach. DTs are very famous when it comes to classification and regression problems. Multiple decision rules are constructed which are inferred from various features of data. They are limited by the fact that they can be very non-robust. A small change in the training data can result in a large change in the tree and consequently the final predictions \cite{james2013introduction}.
The problem of learning an optimal decision tree is known to be NP-complete under several aspects of optimality and even for simple concepts \cite{laurent1976constructing}.
\subsection{Bagging}
Bagging also known as Bootstrap Aggregating \cite{lee2020bootstrap} is a technique in machine learning that involves combining multiple models trained on different subsets of the training data to improve predictive performance. The bagging technique works by creating multiple bootstrap samples of the training data and training a different model on each sample. By averaging the predictions of all the individual models, bagging can reduce overfitting and improve the stability and accuracy of the final prediction. Bagging can be applied to various machine learning algorithms, such as decision trees, neural networks, and random forests.
The main benefits of bagging are improved accuracy, stability, and robustness, making it a popular technique in ensemble learning. Bagging is not always effective with data that has a high degree of correlation or has a very small number of informative features and may not always improve the performance of certain machine learning algorithms, such as k-nearest neighbors, that are inherently stable.
\subsection{Gradient Boosting} 
Gradient boosting \cite{zhang2019predictive, bentejac2021comparative} is a machine learning technique used in regression and classification tasks, among others. It gives a prediction model in the form of an ensemble of weak prediction models, which are typically decision trees. While boosting can increase the accuracy of a base learner, such as a decision tree or linear regression, it sacrifices intelligibility and interpretability.For example, following the path that a decision tree takes to make its decision is trivial and self-explained, but following the paths of hundreds or thousands of trees is much harder.
\subsection{AdaBoost}
AdaBoost, short for Adaptive Boosting \cite{shahraki2020boosting, wang2021improved}, is a statistical classification meta-algorithm, Every learning algorithm tends to suit some problem types better than others, and typically has many different parameters and configurations to adjust before it achieves optimal performance on a dataset. AdaBoost (with decision trees as the weak learners) is often referred to as the best out-of-the-box classifier \cite{kegl2013return}. When used with decision tree learning, information gathered at each stage of the AdaBoost algorithm about the relative 'hardness' of each training sample is fed into the tree growing algorithm such that later trees tend to focus on harder-to-classify examples.  AdaBoost is particularly prone to overfitting on noisy datasets.
\subsection{Support Vector Machine (SVM)}
Support Vector Machine \cite{kurani2023comprehensive, vos2022vibration, koklu2022cnn} is a supervised learning method, which is used for problems such as classifying different classes (classification), predicting continuous value (Regression) and detection of any outliers. SVM is emplyed here because it is very effective in high dimensional spaces, and our dataset has deep feature space. Another reason for using SVM is that it is very memory efficient as it uses a subset of training samples in decision function. When devising the architecture of SVM approach there were 3 primary parameters of concern, Kernel function, gamma, and C. Gamma dictates how much influence can a single sample in training space has. A lower value of C indicates the decision surface of the classifier to be smooth, which means that there can some percentage of mis-classification allowed. However, if C is set to a higher value, then SVM aims to classify all the training samples correctly, and percentage of error or mis-classification is reduced.  Different values of these that were evaluated are given in table \ref{SVM}.
\begin{table}
\caption{SVM model variations}
\label{SVM}
\centering
\setlength{\tabcolsep}{4pt}
\begin{tabular}{|M{85pt}|M{60pt}|M{60pt}|}
\hline
\textbf{Kernel}&\textbf{C-Parameter} & \textbf{Gamma}\\
\hline
\multirow{3}{4em}{\textbf{Linear}} & 1 & 0.1 \\
& 5 & 0.1\\
& 10 & 0.1  \\
\hline
\multirow{3}{4em}{\textbf{Radial Basis Function}}  & 
1 & 0.1 \\
& 5 & 0.1\\
& 10 & 0.1  \\
\hline
\end{tabular}
\end{table}

\subsection{Random Forest (RF)}

When multiple decision trees are combined, and are used as an ensemble strategy to improve the accuracy, this architecture is known as Random Forest \cite{balyan2022hybrid, wang2023risk}. Key parameter for Random forest is the $n\_estimators$ which indicates the number of trees to be used in forest. Another vital parameter is the $max\_depth$ of the trees in forest which limits the number of splits/divisions that can be performed per tree.
Table \ref{tab:RF} gives parameters applied during training using RF. 
\begin{table}
\caption{RF model variations}
\label{tab:RF}
\centering
\setlength{\tabcolsep}{4pt}
\begin{tabular}{|M{120pt}|M{80pt}|}
\hline
\textbf{n-estimators}&\textbf{Depth of trees}\\
\hline
\multirow{3}{4em}{\textbf{100}} & 10 \\
& 20\\
& 30\\
& 40\\
\hline
\multirow{3}{4em}{\textbf{200}}  & 
10 \\
& 20\\
\hline
\end{tabular}
\end{table}

\subsection{XGBoost}
XGBoost \cite{velarde2023evaluating} (eXtreme Gradient Boosting) is an open-source software library which provides a regularizing gradient boosting. Salient features of XGBoost which make it different from other gradient boosting algorithms are clever penalization of trees, proportional shrinking of leaf nodes, Newton Boosting, extra randomization parameter, implementation on single, distributed systems and out-of-core computation and automatic feature selection. It is known for its superior performance in various machine learning tasks\cite{zhang2018data, chen2019xgboost, jiang2019pedestrian}, including malware analysis, for the following reasons: it effectively handles complex relationships and captures intricate patterns within the data, which is essential for accurately detecting and classifying obfuscated malware. XGBoost employs regularization techniques to prevent overfitting and build robust models that generalize well to unseen malware samples. Built on the gradient boosting framework, it learns from previous models' mistakes and leverages the strengths of multiple decision trees to achieve high accuracy. XGBoost includes strategies to handle class imbalance, ensuring effective learning from imbalanced data. Additionally, it offers computational efficiency, scalability, and the ability to handle large-scale analysis efficiently, crucial for processing extensive malware datasets.

\subsection{Voting}
Voting ensemble \cite{hussain2023emotion, zhang2023weighted, sevim2023improving, mohammadifar2023stacking} is a machine learning technique that combines multiple models trained on the same dataset to improve the overall predictive power of the system. In voting ensemble, each model is given an equal vote, and the final prediction is based on the majority vote of all the models.
By combining multiple models trained on the same dataset, voting ensemble can often achieve higher accuracy than any individual model but this leads to complexity and makes it computationally expensive than individual models, as it requires training and combining multiple models. It can reduce the risk of overfitting, as it combines multiple models with different biases and strengths, which helps to reduce the variance in the final predictions and is often more robust than individual models, as it can handle missing or noisy data more effectively by combining the predictions of multiple models. But it may not be effective if the individual models are too similar, as it can lead to over-reliance on certain features or biases and most importantly requires training multiple models, which can be time-consuming and resource-intensive.
\section{Experimental Environment}
Hardware and software resources employed during experiments are given in table \ref{Hardware_Resources} and \ref{Software_Resources}, respectively.
\begin{table}
\caption{Hardware Resources}
\label{Hardware_Resources}
\centering
\setlength{\tabcolsep}{3pt}
\begin{tabular}{|M{120pt}|M{100pt}|}
\hline
Name of hardware & Specification \\
\hline
Intel(R) Core(TM) i7-8700 &	CPU @ 3.20GHz\\\hline
RAM &	32GB\\\hline
\end{tabular}
\end{table}
\begin{table}
\caption{Software Resources}
\label{Software_Resources}
\centering
\setlength{\tabcolsep}{3pt}
\begin{tabular}{|M{70pt}|M{85pt}|M{300pt}|}
\hline
Name of software &Source & Description \\
\hline
Python3.9.15&	www.python.org &	Platform independent programming language (open source)\\\hline
TensorFlow2.10.0 &	www.tensorflow.org/&	End-to-end learning framework for deploying machine learning models (open source)\\\hline
Pytorch1.13.1	&	www.pytorch.org& Large-scale deep machine learning library (open source)\\\hline
Scikit-learn & www.scikit-learn.org/stable/& Simple open-source efficient predictive data analysis tool\\\hline
Microsoft Windows 11 &  www.microsoft.com/en-us/windows/?r=1 & The most recent major version of Microsoft's Windows NT operating system\\\hline
\end{tabular}
\end{table}
\section{Results and Discussion}
\label{sec:Results}

\begin{table}
\caption{Results}
\label{Results}
\centering
\setlength{\tabcolsep}{3pt}
\begin{tabular}{|M{85pt}|M{37pt}|M{37pt}|M{47pt}|M{37pt}|}
\hline
Classifier(s)&Precision& Recall & F-1 Score & Accuracy \\\hline
Decision Tree&	74&	76&	72&75\\\hline
Logistic Regression&	82&	72&	69&	76\\\hline
Random Forest&91&	77&	79&	82\\\hline
XGBoost	&88&	80	&82	&83\\\hline
SVM	&83&	71&	76&	73\\\hline
AdaBoostClassifier&	88&	76&	76&	80\\\hline
Ensemble (Voting:SVM+LR) &	79&	77&	77&	77\\\hline
\end{tabular}
\end{table}
\begin{figure}\centering
\subfloat{{\includegraphics[width=9cm]{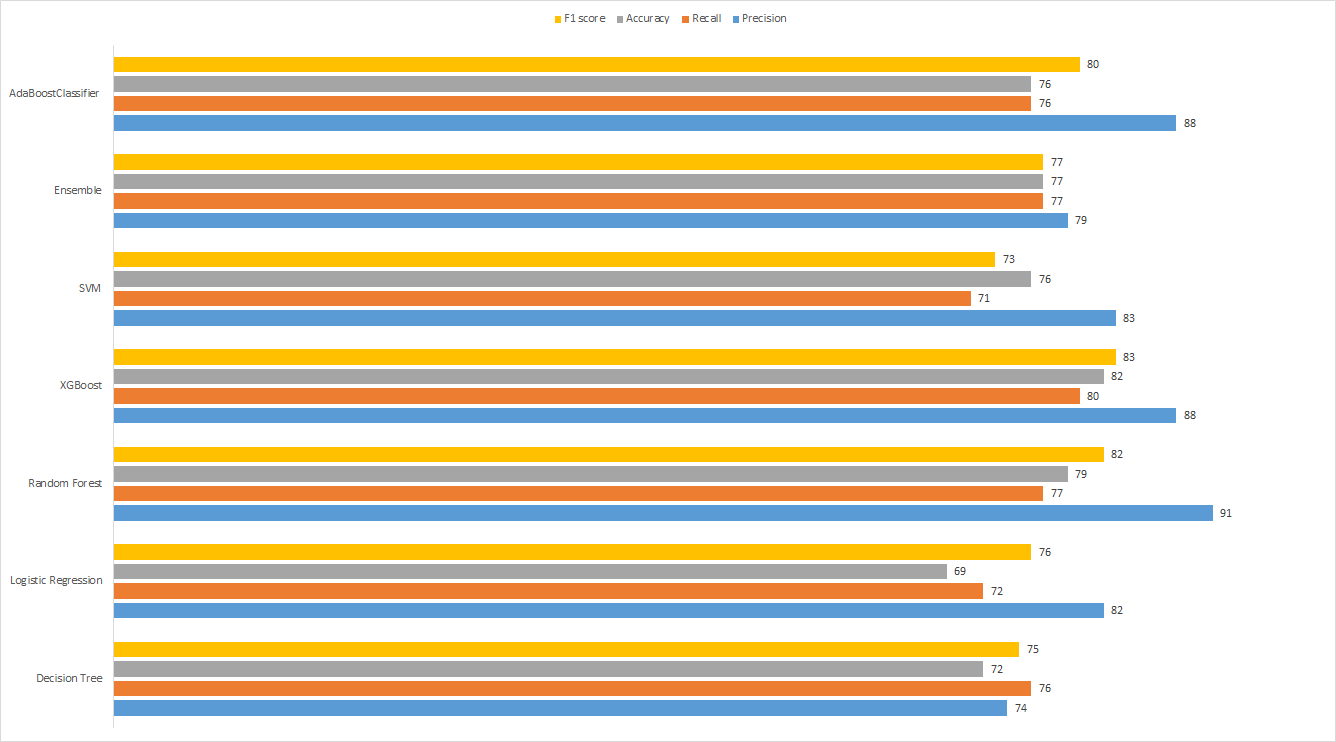} }}%
    \caption{Comparison between Decision Trees, AdaBoost,  SVM, RF, Logistic Regression, XGBoost, and Ensemble of SVM and LR using Recall, Accuracy, Precision and F-Score, XGBoost and RF are best performing ones}
    \label{Results}
    \end{figure}
    \begin{figure}\centering
\subfloat{{\includegraphics[width=9cm]{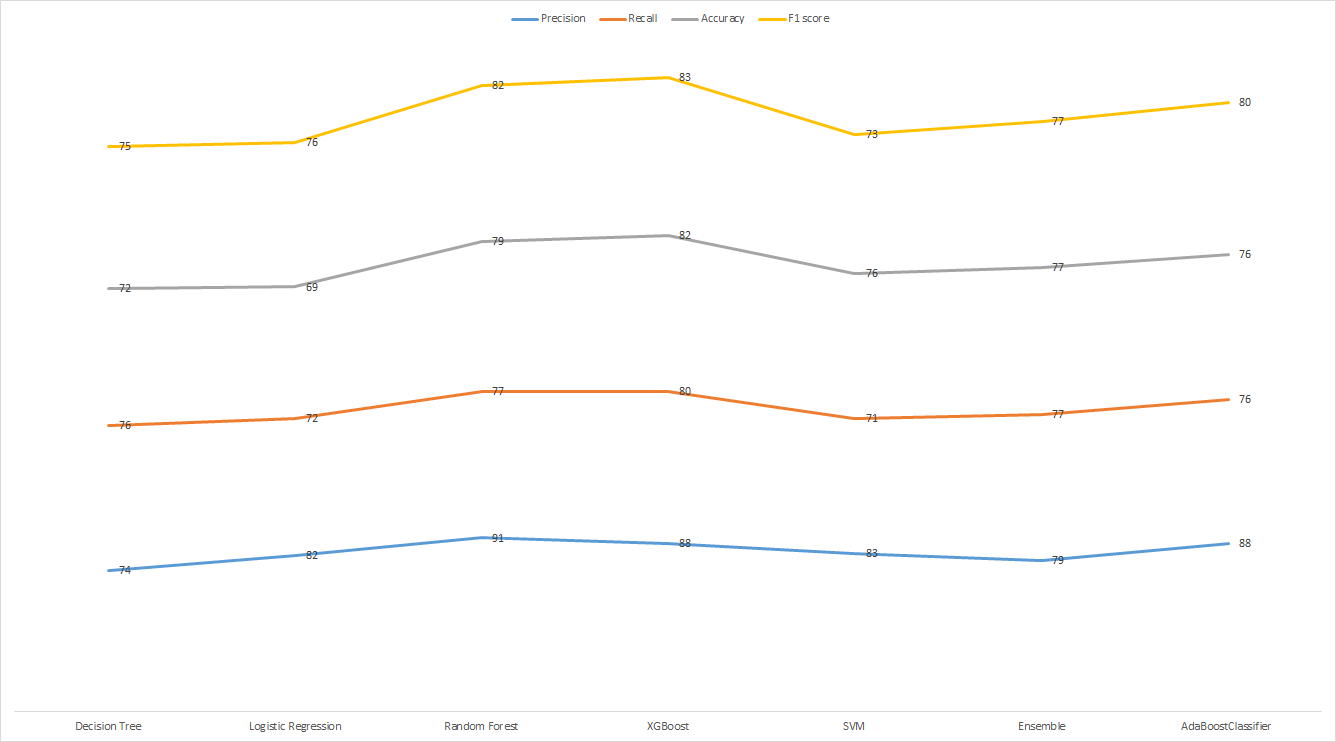} }}%
    \caption{XGBoost and RF perform best with RF edging it on precision while in other three metrics XGBoost performs better}
    \label{Results2}
    \end{figure}
Results demonstrate that XGBoost outperforms all other methods achieving an accuracy, precision, recall, F1-Score of 82\%, 88\%, 80\% and 83\% respectively as shown in fig. \ref{Results}. RF has the highest precision shown in fig. \ref{Results2} at 88\% but in other three metrics XGBoost performs better and overall gives a better result.
\section{Conclusion}
With the rise in computer usage, cybersecurity has emerged as a crucial concern in the digital era. Cybercriminals have expanded their activities beyond traditional hacking and virus distribution. Daily malware attacks inflict significant financial losses by targeting computer users, businesses, and government agencies. Despite the availability of diverse security tools, malware can evade detection by making clever adjustments, causing significant challenges for security experts. To address this issue, the Obfuscated Malware Dataset (OMD) is introduced, consisting of 21,924 samples from 40 distinct malware families. The dataset incorporates various obfuscation techniques, simulating the tactics employed by malware authors to create new strains. Robust models utilizing traditional machine learning algorithms, such as SVM, RF, XGBOOST, and others, are trained to effectively identify these evasive malware instances that are difficult to detect through conventional means.   
\bibliography{OMD}

\begin{thebibliography}{52}
\providecommand{\natexlab}[1]{#1}
\providecommand{\url}[1]{\texttt{#1}}
\expandafter\ifx\csname urlstyle\endcsname\relax
  \providecommand{\doi}[1]{doi: #1}\else
  \providecommand{\doi}{doi: \begingroup \urlstyle{rm}\Url}\fi

\bibitem[Gao et~al.(2019)Gao, Li, Kong, Bissyand{\'e}, and Klein]{gao2019should}
Jun Gao, Li~Li, Pingfan Kong, Tegawend{\'e}~F Bissyand{\'e}, and Jacques Klein.
\newblock Should you consider adware as malware in your study?
\newblock In \emph{2019 IEEE 26th International Conference on Software Analysis, Evolution and Reengineering (SANER)}, pages 604--608. IEEE, 2019.

\bibitem[Kumar and Subbiah(2022)]{kumar2022zero}
Rajesh Kumar and Geetha Subbiah.
\newblock Zero-day malware detection and effective malware analysis using shapley ensemble boosting and bagging approach.
\newblock \emph{Sensors}, 22\penalty0 (7):\penalty0 2798, 2022.

\bibitem[Barros et~al.(2022)Barros, Chagas, Oliveira, Queiroz, and Ramos]{barros2022malware}
Pedro~H Barros, Eduarda~TC Chagas, Leonardo~B Oliveira, Fabiane Queiroz, and Heitor~S Ramos.
\newblock Malware-smell: A zero-shot learning strategy for detecting zero-day vulnerabilities.
\newblock \emph{Computers \& Security}, 120:\penalty0 102785, 2022.

\bibitem[Botacin et~al.(2022)Botacin, Alves, Oliveira, and Gr{\'e}gio]{botacin2022heaven}
Marcus Botacin, Marco~Zanata Alves, Daniela Oliveira, and Andr{\'e} Gr{\'e}gio.
\newblock Heaven: A hardware-enhanced antivirus engine to accelerate real-time, signature-based malware detection.
\newblock \emph{Expert Systems with Applications}, 201:\penalty0 117083, 2022.

\bibitem[Selamat and Ali(2022)]{selamat2022polymorphic}
Nur~Syuhada Selamat and Fakariah Hani~Mohd Ali.
\newblock Polymorphic malware detection based on supervised machine learning.
\newblock \emph{Journal of Positive School Psychology}, 6\penalty0 (3):\penalty0 8538--8547, 2022.

\bibitem[Brewer(2016)]{brewer2016ransomware}
Ross Brewer.
\newblock Ransomware attacks: detection, prevention and cure.
\newblock \emph{Network Security}, 2016\penalty0 (9):\penalty0 5--9, 2016.

\bibitem[Tahir(2018)]{tahir2018study}
Rabia Tahir.
\newblock A study on malware and malware detection techniques.
\newblock \emph{International Journal of Education and Management Engineering}, 8\penalty0 (2):\penalty0 20, 2018.

\bibitem[Menn(2010)]{menn2010fatal}
Joseph Menn.
\newblock \emph{Fatal system error: the hunt for the new crime lords who are bringing down the internet}.
\newblock PublicAffairs, 2010.

\bibitem[Geer and Conway(2009)]{geer20090wned}
Daniel~E Geer and Daniel~G Conway.
\newblock the 0wned price index.
\newblock \emph{IEEE Security \& Privacy}, 7\penalty0 (1):\penalty0 86--87, 2009.

\bibitem[Salloum et~al.(2022)Salloum, Gaber, Vadera, and Sharan]{salloum2022systematic}
Said Salloum, Tarek Gaber, Sunil Vadera, and Khaled Sharan.
\newblock A systematic literature review on phishing email detection using natural language processing techniques.
\newblock \emph{IEEE Access}, 2022.

\bibitem[Wang et~al.(2021)Wang, Long, Wang, Liu, and Fu]{wang2021deep}
Haojun Wang, Haixia Long, Ailan Wang, Tianyue Liu, and Haiyan Fu.
\newblock Deep learning and regularization algorithms for malicious code classification.
\newblock \emph{IEEE Access}, 9:\penalty0 91512--91523, 2021.

\bibitem[Grammatikakis et~al.(2021)Grammatikakis, Koufos, Kolokotronis, Vassilakis, and Shiaeles]{grammatikakis2021understanding}
Konstantinos~P Grammatikakis, Ioannis Koufos, Nicholas Kolokotronis, Costas Vassilakis, and Stavros Shiaeles.
\newblock Understanding and mitigating banking trojans: From zeus to emotet.
\newblock In \emph{2021 IEEE International Conference on Cyber Security and Resilience (CSR)}, pages 121--128. IEEE, 2021.

\bibitem[Kazi et~al.(2022)Kazi, Woodhead, and Gan]{kazi2022comparing}
Mohamed~Ali Kazi, Steve Woodhead, and Diane Gan.
\newblock Comparing the performance of supervised machine learning algorithms when used with a manual feature selection process to detect zeus malware.
\newblock \emph{International Journal of Grid and Utility Computing}, 13\penalty0 (5):\penalty0 495--504, 2022.

\bibitem[Rose et~al.(2022)Rose, Swann, Grammatikakis, Koufos, Bendiab, Shiaeles, and Kolokotronis]{rose2022ideres}
Joseph~R Rose, Matthew Swann, Konstantinos~P Grammatikakis, Ioannis Koufos, Gueltoum Bendiab, Stavros Shiaeles, and Nicholas Kolokotronis.
\newblock Ideres: Intrusion detection and response system using machine learning and attack graphs.
\newblock \emph{Journal of Systems Architecture}, 131:\penalty0 102722, 2022.

\bibitem[Nataraj et~al.(2011)Nataraj, Karthikeyan, Jacob, and Manjunath]{nataraj2011malware}
Lakshmanan Nataraj, Sreejith Karthikeyan, Gregoire Jacob, and Bangalore~S Manjunath.
\newblock Malware images: visualization and automatic classification.
\newblock In \emph{Proceedings of the 8th international symposium on visualization for cyber security}, pages 1--7, 2011.

\bibitem[Gibert et~al.(2019)Gibert, Mateu, Planes, and Vicens]{gibert2019using}
Daniel Gibert, Carles Mateu, Jordi Planes, and Ramon Vicens.
\newblock Using convolutional neural networks for classification of malware represented as images.
\newblock \emph{Journal of Computer Virology and Hacking Techniques}, 15:\penalty0 15--28, 2019.

\bibitem[Ronen et~al.(2018)Ronen, Radu, Feuerstein, Yom-Tov, and Ahmadi]{ronen2018microsoft}
Royi Ronen, Marian Radu, Corina Feuerstein, Elad Yom-Tov, and Mansour Ahmadi.
\newblock Microsoft malware classification challenge.
\newblock \emph{arXiv preprint arXiv:1802.10135}, 2018.

\bibitem[Kancherla and Mukkamala(2013)]{kancherla2013image}
Kesav Kancherla and Srinivas Mukkamala.
\newblock Image visualization based malware detection.
\newblock In \emph{2013 IEEE Symposium on Computational Intelligence in Cyber Security (CICS)}, pages 40--44. IEEE, 2013.

\bibitem[Ahmadi et~al.(2016)Ahmadi, Ulyanov, Semenov, Trofimov, and Giacinto]{ahmadi2016novel}
Mansour Ahmadi, Dmitry Ulyanov, Stanislav Semenov, Mikhail Trofimov, and Giorgio Giacinto.
\newblock Novel feature extraction, selection and fusion for effective malware family classification.
\newblock In \emph{Proceedings of the sixth ACM conference on data and application security and privacy}, pages 183--194, 2016.

\bibitem[Rezende et~al.(2017)Rezende, Ruppert, Carvalho, Ramos, and De~Geus]{rezende2017malicious}
Edmar Rezende, Guilherme Ruppert, Tiago Carvalho, Fabio Ramos, and Paulo De~Geus.
\newblock Malicious software classification using transfer learning of resnet-50 deep neural network.
\newblock In \emph{2017 16th IEEE International Conference on Machine Learning and Applications (ICMLA)}, pages 1011--1014. IEEE, 2017.

\bibitem[Qiao et~al.(2016)Qiao, Yun, and Zhang]{qiao2016automatically}
Yanchen Qiao, Xiaochun Yun, and Yongzheng Zhang.
\newblock How to automatically identify the homology of different malware.
\newblock In \emph{2016 IEEE Trustcom/BigDataSE/ISPA}, pages 929--936. IEEE, 2016.

\bibitem[Karbab et~al.(2018)Karbab, Debbabi, Derhab, and Mouheb]{karbab2018maldozer}
ElMouatez~Billah Karbab, Mourad Debbabi, Abdelouahid Derhab, and Djedjiga Mouheb.
\newblock Maldozer: Automatic framework for android malware detection using deep learning.
\newblock \emph{Digital Investigation}, 24:\penalty0 S48--S59, 2018.

\bibitem[Arp et~al.(2014)Arp, Spreitzenbarth, Hubner, Gascon, Rieck, and Siemens]{arp2014drebin}
Daniel Arp, Michael Spreitzenbarth, Malte Hubner, Hugo Gascon, Konrad Rieck, and CERT Siemens.
\newblock Drebin: Effective and explainable detection of android malware in your pocket.
\newblock In \emph{Ndss}, volume~14, pages 23--26, 2014.

\bibitem[Anderson and Roth(2018)]{anderson2018ember}
Hyrum~S Anderson and Phil Roth.
\newblock Ember: an open dataset for training static pe malware machine learning models.
\newblock \emph{arXiv preprint arXiv:1804.04637}, 2018.

\bibitem[Sebasti{\'a}n et~al.(2016)Sebasti{\'a}n, Rivera, Kotzias, and Caballero]{sebastian2016avclass}
Marcos Sebasti{\'a}n, Richard Rivera, Platon Kotzias, and Juan Caballero.
\newblock Avclass: A tool for massive malware labeling.
\newblock In \emph{Research in Attacks, Intrusions, and Defenses: 19th International Symposium, RAID 2016, Paris, France, September 19-21, 2016, Proceedings 19}, pages 230--253. Springer, 2016.

\bibitem[Plohmann et~al.(2017)Plohmann, Clauss, Enders, and Padilla]{plohmann2017malpedia}
Daniel Plohmann, Martin Clauss, Steffen Enders, and Elmar Padilla.
\newblock Malpedia: a collaborative effort to inventorize the malware landscape.
\newblock \emph{Proceedings of the Botconf}, 2017.

\bibitem[Kotzias et~al.(2015)Kotzias, Matic, Rivera, and Caballero]{kotzias2015certified}
Platon Kotzias, Srdjan Matic, Richard Rivera, and Juan Caballero.
\newblock Certified pup: abuse in authenticode code signing.
\newblock In \emph{Proceedings of the 22nd ACM SIGSAC Conference on Computer and Communications Security}, pages 465--478, 2015.

\bibitem[Huang and Stokes(2016)]{huang2016mtnet}
Wenyi Huang and Jack~W Stokes.
\newblock Mtnet: a multi-task neural network for dynamic malware classification.
\newblock In \emph{Detection of Intrusions and Malware, and Vulnerability Assessment: 13th International Conference, DIMVA 2016, San Sebasti{\'a}n, Spain, July 7-8, 2016, Proceedings 13}, pages 399--418. Springer, 2016.

\bibitem[Shorten and Khoshgoftaar(2019)]{shorten2019survey}
Connor Shorten and Taghi~M Khoshgoftaar.
\newblock A survey on image data augmentation for deep learning.
\newblock \emph{Journal of big data}, 6\penalty0 (1):\penalty0 1--48, 2019.

\bibitem[Naveed et~al.(2021)Naveed, Anwar, Hayat, Javed, and Mian]{naveed2021survey}
Humza Naveed, Saeed Anwar, Munawar Hayat, Kashif Javed, and Ajmal Mian.
\newblock Survey: Image mixing and deleting for data augmentation.
\newblock \emph{arXiv preprint arXiv:2106.07085}, 2021.

\bibitem[Charbuty and Abdulazeez(2021)]{charbuty2021classification}
Bahzad Charbuty and Adnan Abdulazeez.
\newblock Classification based on decision tree algorithm for machine learning.
\newblock \emph{Journal of Applied Science and Technology Trends}, 2\penalty0 (01):\penalty0 20--28, 2021.

\bibitem[James et~al.(2013)James, Witten, Hastie, and Tibshirani]{james2013introduction}
Gareth James, Daniela Witten, Trevor Hastie, and Robert Tibshirani.
\newblock \emph{An introduction to statistical learning}, volume 112.
\newblock Springer, 2013.

\bibitem[Laurent and Rivest(1976)]{laurent1976constructing}
Hyafil Laurent and Ronald~L Rivest.
\newblock Constructing optimal binary decision trees is np-complete.
\newblock \emph{Information processing letters}, 5\penalty0 (1):\penalty0 15--17, 1976.

\bibitem[Lee et~al.(2020)Lee, Ullah, and Wang]{lee2020bootstrap}
Tae-Hwy Lee, Aman Ullah, and Ran Wang.
\newblock Bootstrap aggregating and random forest.
\newblock \emph{Macroeconomic forecasting in the era of big data: Theory and practice}, pages 389--429, 2020.

\bibitem[Zhang et~al.(2019)Zhang, Zhao, Canes, Steinberg, Lyashevska, et~al.]{zhang2019predictive}
Zhongheng Zhang, Yiming Zhao, Aran Canes, Dan Steinberg, Olga Lyashevska, et~al.
\newblock Predictive analytics with gradient boosting in clinical medicine.
\newblock \emph{Annals of translational medicine}, 7\penalty0 (7), 2019.

\bibitem[Bent{\'e}jac et~al.(2021)Bent{\'e}jac, Cs{\"o}rg{\H{o}}, and Mart{\'\i}nez-Mu{\~n}oz]{bentejac2021comparative}
Candice Bent{\'e}jac, Anna Cs{\"o}rg{\H{o}}, and Gonzalo Mart{\'\i}nez-Mu{\~n}oz.
\newblock A comparative analysis of gradient boosting algorithms.
\newblock \emph{Artificial Intelligence Review}, 54:\penalty0 1937--1967, 2021.

\bibitem[Shahraki et~al.(2020)Shahraki, Abbasi, and Haugen]{shahraki2020boosting}
Amin Shahraki, Mahmoud Abbasi, and {\O}ystein Haugen.
\newblock Boosting algorithms for network intrusion detection: A comparative evaluation of real adaboost, gentle adaboost and modest adaboost.
\newblock \emph{Engineering Applications of Artificial Intelligence}, 94:\penalty0 103770, 2020.

\bibitem[Wang and Sun(2021)]{wang2021improved}
Wenyang Wang and Dongchu Sun.
\newblock The improved adaboost algorithms for imbalanced data classification.
\newblock \emph{Information Sciences}, 563:\penalty0 358--374, 2021.

\bibitem[K{\'e}gl(2013)]{kegl2013return}
Bal{\'a}zs K{\'e}gl.
\newblock The return of adaboost. mh: multi-class hamming trees.
\newblock \emph{arXiv preprint arXiv:1312.6086}, 2013.

\bibitem[Kurani et~al.(2023)Kurani, Doshi, Vakharia, and Shah]{kurani2023comprehensive}
Akshit Kurani, Pavan Doshi, Aarya Vakharia, and Manan Shah.
\newblock A comprehensive comparative study of artificial neural network (ann) and support vector machines (svm) on stock forecasting.
\newblock \emph{Annals of Data Science}, 10\penalty0 (1):\penalty0 183--208, 2023.

\bibitem[Vos et~al.(2022)Vos, Peng, Jenkins, Shahriar, Borghesani, and Wang]{vos2022vibration}
Kilian Vos, Zhongxiao Peng, Christopher Jenkins, Md~Rifat Shahriar, Pietro Borghesani, and Wenyi Wang.
\newblock Vibration-based anomaly detection using lstm/svm approaches.
\newblock \emph{Mechanical Systems and Signal Processing}, 169:\penalty0 108752, 2022.

\bibitem[Koklu et~al.(2022)Koklu, Unlersen, Ozkan, Aslan, and Sabanci]{koklu2022cnn}
Murat Koklu, M~Fahri Unlersen, Ilker~Ali Ozkan, M~Fatih Aslan, and Kadir Sabanci.
\newblock A cnn-svm study based on selected deep features for grapevine leaves classification.
\newblock \emph{Measurement}, 188:\penalty0 110425, 2022.

\bibitem[Balyan et~al.(2022)Balyan, Ahuja, Lilhore, Sharma, Manoharan, Algarni, Elmannai, and Raahemifar]{balyan2022hybrid}
Amit~Kumar Balyan, Sachin Ahuja, Umesh~Kumar Lilhore, Sanjeev~Kumar Sharma, Poongodi Manoharan, Abeer~D Algarni, Hela Elmannai, and Kaamran Raahemifar.
\newblock A hybrid intrusion detection model using ega-pso and improved random forest method.
\newblock \emph{Sensors}, 22\penalty0 (16):\penalty0 5986, 2022.

\bibitem[Wang et~al.(2023)Wang, Rao, Goh, and Xiao]{wang2023risk}
Jing Wang, Congjun Rao, Mark Goh, and Xinping Xiao.
\newblock Risk assessment of coronary heart disease based on cloud-random forest.
\newblock \emph{Artificial Intelligence Review}, 56\penalty0 (1):\penalty0 203--232, 2023.

\bibitem[Velarde et~al.(2023)Velarde, Sudhir, Deshmane, Deshmunkh, Sharma, and Joshi]{velarde2023evaluating}
Gissel Velarde, Anindya Sudhir, Sanjay Deshmane, Anuj Deshmunkh, Khushboo Sharma, and Vaibhav Joshi.
\newblock Evaluating xgboost for balanced and imbalanced data: Application to fraud detection.
\newblock \emph{arXiv preprint arXiv:2303.15218}, 2023.

\bibitem[Zhang et~al.(2018)Zhang, Qian, Mao, Huang, Huang, and Si]{zhang2018data}
Dahai Zhang, Liyang Qian, Baijin Mao, Can Huang, Bin Huang, and Yulin Si.
\newblock A data-driven design for fault detection of wind turbines using random forests and xgboost.
\newblock \emph{Ieee Access}, 6:\penalty0 21020--21031, 2018.

\bibitem[Chen et~al.(2019)Chen, Liu, Chen, Liu, Zhang, and Liu]{chen2019xgboost}
Minghua Chen, Qunying Liu, Shuheng Chen, Yicen Liu, Chang-Hua Zhang, and Ruihua Liu.
\newblock Xgboost-based algorithm interpretation and application on post-fault transient stability status prediction of power system.
\newblock \emph{IEEE Access}, 7:\penalty0 13149--13158, 2019.

\bibitem[Jiang et~al.(2019)Jiang, Tong, Yin, and Xiong]{jiang2019pedestrian}
Yu~Jiang, Guoxiang Tong, Henan Yin, and Naixue Xiong.
\newblock A pedestrian detection method based on genetic algorithm for optimize xgboost training parameters.
\newblock \emph{IEEE Access}, 7:\penalty0 118310--118321, 2019.

\bibitem[Hussain et~al.(2023)Hussain, AboAlSamh, Ullah, et~al.]{hussain2023emotion}
Muhammad Hussain, Hatim~A AboAlSamh, Ihsan Ullah, et~al.
\newblock Emotion recognition system based on two-level ensemble of deep-convolutional neural network models.
\newblock \emph{IEEE Access}, 11:\penalty0 16875--16895, 2023.

\bibitem[Zhang et~al.(2023)Zhang, Zheng, and Zou]{zhang2023weighted}
Tie Zhang, Jingfu Zheng, and Yanbiao Zou.
\newblock Weighted voting ensemble method for predicting workpiece imaging dimensional deviation based on monocular vision systems.
\newblock \emph{Optics \& Laser Technology}, 159:\penalty0 109012, 2023.

\bibitem[Sevim et~al.(2023)Sevim, Omurca, and Ekinci]{sevim2023improving}
Semih Sevim, Sevin{\c{c}}~{\.I}lhan Omurca, and Ekin Ekinci.
\newblock Improving accuracy of document image classification through soft voting ensemble.
\newblock In \emph{Smart Applications with Advanced Machine Learning and Human-Centred Problem Design}, pages 161--173. Springer, 2023.

\bibitem[Mohammadifar et~al.(2023)Mohammadifar, Gholami, and Golzari]{mohammadifar2023stacking}
Aliakbar Mohammadifar, Hamid Gholami, and Shahram Golzari.
\newblock Stacking-and voting-based ensemble deep learning models (sedl and vedl) and active learning (al) for mapping land subsidence.
\newblock \emph{Environmental Science and Pollution Research}, 30\penalty0 (10):\penalty0 26580--26595, 2023.

\end{thebibliography}
    


\bibliographystyle{unsrtnat}






\end{document}